\documentclass [12pt,a4paper]{article}
\usepackage[hyperindex,breaklinks]{hyperref}
\usepackage{supertabular}
\newcommand{\E}{\mathrm{E}}
\newcommand{\N}{\phantom{\mathrm{E}}}

\title{A list of peculiar velocities and distances to 1623 galaxies from the
Revised Flat Galaxy Catalogue}

\author{S.L. Parnovsky \thanks{Astronomical Observatory of Kyiv Taras
Shevchenko National University, Observatorna str. 3, 04053, Kyiv, Ukraine}
\and A.S. Parnowski \thanks{Space Research Institute of National Academy of
Sciences of Ukraine and National Space Agency of Ukraine, Acad. Glushkova
prosp. 40, korp. 4/1, 03680 MSP, Kyiv-187, Ukraine}}

\begin{document}
\maketitle
\begin{abstract}
We present a list of distances and peculiar velocities for 1623 RFGC galaxies
for three models of collective large-scale galaxy motion on distances about
$100 h^{-1}$ Mpc. It is based upon the article [arXiv:0910.4640].

The ASCII version of the list can be downloaded from the AO KNU website:
http://www.observ.univ.kiev.ua/data/rfgcvpec.zip
\end{abstract}

\section{Introduction}
In a recent article \cite{1} the parameters of collective large-scale motion
of flat spiral galaxies from catalogues \cite{2} and \cite{3} were found. The
data about the HI line widths and redshifts for 1623 galaxies were used. We
used the generalized Tully-Fisher relation in the ``HI line width -- linear
diameter'' version. Here we present distances to these galaxies and their
peculiar velocities calculated on the basis of these results. Please refer to
the original article \cite{1} for information on data and models used.

This list is an extension of a similar list obtained in the article \cite{4}
and updated later in the article \cite{5}. Since the results in the paper
\cite{1} were obtained for three models, here we also present the data for
three models. These three models are: Hubble expansion + dipole (D-model),
Hubble expansion + dipole + quadrupole (DQ-model), and Hubble expansion +
dipole + quadrupole + octopole (DQO-model). The distance to galaxy is
calculated using the following values: corrected red and blue angular
diameters, axial ratio, morphological type, surface brightness index, and the
corrected HI line width. All the formulae and the numerical values of the
coefficients are given in the paper \cite{1}. For calculations we used the
values for the subsample with upper distance limit of $100 h^{-1}$ Mpc.

\section{Description of tables}
Due to the large number of columns, the list was split into two tables. The
first one contains, in order:
\begin{enumerate}
\item{RFGC number}
\item{FGC/FGCE number}
\item{Right Ascension and Declination for epoch J2000.0}
\item{Corrected blue (POSS-I) diameter $a_b$, arcmin}
\item{Corrected red (ESO/SERC) diameter $a_r$, arcmin}
\item{HI line width $W_{50}$, km/s}
\item{Reshift velocity in the CMB frame $V_{3K}$, km/s}
\item{Distance to a galaxy in D-model, km/s}
\item{Dipole component of the velocity in D-model $V^{dip}$, km/s}
\item{Regression velocity (Hubble expansion + multipole) in D-model $V^{reg}$,
km/s}
\item{Deviation of regression velocity from redshift velocity in D-model
$\delta V$, km/s}
\item{Peculiar velocity in D-model $V^{pec}$, km/s}
\end{enumerate}
The second one contains, in order:
\begin{enumerate}
\item{RFGC number (the same as in the first table)}
\item{Distance to a galaxy in DQ-model, km/s}
\item{Dipole component of the velocity in DQ-model $V^{dip}$, km/s}
\item{Quadrupole component of the velocity in DQ-model $V^{qua}$, km/s}
\item{Regression velocity (Hubble expansion + multipole) in DQ-model
$V^{reg}$, km/s}
\item{Deviation of regression velocity from redshift velocity in DQ-model
$\delta V$, km/s}
\item{Peculiar velocity in DQ-model $V^{pec}$, km/s}
\item{Distance to a galaxy in DQO-model, km/s}
\item{Dipole component of the velocity in DQO-model $V^{dip}$, km/s}
\item{Quadrupole component of the velocity in DQO-model $V^{qua}$, km/s}
\item{Octopole component of the velocity in DQO-model $V^{oct}$, km/s}
\item{Regression velocity (Hubble expansion + multipole) in DQO-model
$V^{reg}$, km/s}
\item{Deviation of regression velocity from redshift velocity in DQO-model
$\delta V$, km/s}
\item{Peculiar velocity in DQO-model $V^{pec}$, km/s}
\end{enumerate}
Other values and their descriptions are available in the catalogue itself
\cite{3}.

The ASCII version of the list can be downloaded from the AO KNU website:
http://www.observ.univ.kiev.ua/data/rfgcvpec.zip

\addtolength{\textwidth}{1in}
\addtolength{\textheight}{2in}
\clearpage
\addtolength{\hoffset}{-0.4in}
\addtolength{\voffset}{-0.7in}
\setlength{\oddsidemargin}{0pt}
\setlength{\topmargin}{0pt}
\setlength{\footskip}{0pt}

\begin{center}
\topcaption{Main galaxy parameters and a list of velocity and distance data for
the RFGC galaxies in the framework of D-model}
\tablehead{
\hline
\multicolumn{1}{c}{RFGC}&
\multicolumn{1}{c|}{FGC}&
\multicolumn{1}{c}{RA (2000) D}&
\multicolumn{1}{c}{$a_b$}&
\multicolumn{1}{c}{$a_r$}&
\multicolumn{1}{c}{$W_{50}$}&
\multicolumn{1}{c|}{$V_{3K}$}&
\multicolumn{1}{c}{$Hr_d$}&
\multicolumn{1}{c}{$V^{dip}$}&
\multicolumn{1}{c}{$V^{reg}$}&
\multicolumn{1}{c}{$\delta V$}&
\multicolumn{1}{c}{$V^{pec}$} \\ \hline}
\tabletail{\hline}
\begin{supertabular}{r@{~}r|r@{~}r@{~}r@{~}r@{~}r|r@{~}r@{~}r@{~}r@{~}r}\hline
$   1$&$2565\N$&$000056.1+202017$&$1.33$&$1.27$&$416.3$&$ 6452$&$ 6485$&$ -218$&$ 6267$&$  185$&$  -33$\\
$   4$&$2567\N$&$000136.8+033017$&$ .94$&$ .88$&$241.7$&$ 5982$&$ 5557$&$ -194$&$ 5363$&$  620$&$  426$\\
$   6$&$2563\N$&$000204.3+803832$&$ .98$&$ .92$&$215.7$&$ 4011$&$ 4327$&$ -155$&$ 4172$&$ -161$&$ -316$\\
$  12$&$2573\N$&$000223.3+271238$&$1.19$&$1.00$&$277.7$&$ 7291$&$ 5063$&$ -224$&$ 4839$&$ 2452$&$ 2228$\\
$  14$&$   1\N$&$000403.0-065845$&$ .56$&$ .53$&$106.6$&$ 3381$&$ 3116$&$ -171$&$ 2945$&$  437$&$  265$\\
$  22$&$   6\E$&$000541.0-485848$&$ .51$&$ .45$&$120.7$&$ 6411$&$ 4312$&$  -29$&$ 4283$&$ 2128$&$ 2099$\\
$  26$&$   5\N$&$000638.1+171703$&$ .82$&$ .86$&$ 85.7$&$  529$&$ 1527$&$ -219$&$ 1308$&$ -779$&$ -998$\\
$  27$&$   6\N$&$000653.7+414426$&$1.09$&$ .92$&$254.3$&$ 3951$&$ 4986$&$ -226$&$ 4760$&$ -809$&$-1035$\\
\end{supertabular}
\end{center}

\clearpage
\begin{center}
\topcaption{A list of velocity and distance data for the RFGC galaxies in the
framework of DQ- and DQO-models}
\tablehead{
\hline
\multicolumn{1}{c|}{RFGC}&
\multicolumn{1}{c}{$Hr_q$}&
\multicolumn{1}{c}{$V^{dip}$}&
\multicolumn{1}{c}{$V^{qua}$}&
\multicolumn{1}{c}{$V^{reg}$}&
\multicolumn{1}{c}{$\delta V$}&
\multicolumn{1}{c|}{$V^{pec}$}&
\multicolumn{1}{c}{$Hr_o$}&
\multicolumn{1}{c}{$V^{dip}$}&
\multicolumn{1}{c}{$V^{qua}$}&
\multicolumn{1}{c}{$V^{oct}$}&
\multicolumn{1}{c}{$V^{reg}$}&
\multicolumn{1}{c}{$\delta V$}&
\multicolumn{1}{c}{$V^{pec}$}\\ \hline}
\tabletail{\hline}
\begin{supertabular}{r|r@{~}r@{~}r@{~}r@{~}r@{~}r|r@{~}r@{~}r@{~}r@{~}r@{~}r@{~}r}\hline
$   1$&$ 6472$&$ -174$&$  -99$&$ 6199$&$  253$&$  -20$&$ 6448$&$ -190$&$  -59$&$  226$&$ 6425$&$   27$&$    4$\\
$   4$&$ 5566$&$ -139$&$   63$&$ 5490$&$  492$&$  416$&$ 5569$&$ -156$&$  145$&$  216$&$ 5773$&$  210$&$  414$\\
$   6$&$ 4314$&$ -171$&$ -103$&$ 4041$&$  -30$&$ -304$&$ 4288$&$ -172$&$ -101$&$   10$&$ 4026$&$  -15$&$ -277$\\
$  12$&$ 5073$&$ -185$&$ -125$&$ 4763$&$ 2528$&$ 2218$&$ 5042$&$ -200$&$ -109$&$  124$&$ 4857$&$ 2434$&$ 2249$\\
$  14$&$ 3075$&$ -112$&$   85$&$ 3048$&$  333$&$  307$&$ 3045$&$ -128$&$  142$&$   71$&$ 3130$&$  251$&$  337$\\
$  22$&$ 4289$&$   28$&$  195$&$ 4512$&$ 1900$&$ 2123$&$ 4254$&$   16$&$  289$&$  118$&$ 4677$&$ 1734$&$ 2157$\\
$  26$&$ 1498$&$ -172$&$  -14$&$ 1312$&$ -783$&$ -969$&$ 1482$&$ -185$&$   -3$&$   13$&$ 1306$&$ -777$&$ -953$\\
$  27$&$ 4994$&$ -200$&$ -193$&$ 4601$&$ -650$&$-1043$&$ 4960$&$ -210$&$ -202$&$   94$&$ 4642$&$ -691$&$-1009$\\
\end{supertabular}
\end{center}
\end{document}